\documentclass[%
reprint,
superscriptaddress,
 amsmath,amssymb,
 aps,
prl,
longbibliography,
]{revtex4-1}

\usepackage{graphicx}
\usepackage{dcolumn}
\usepackage{bm}

\usepackage{color}
\usepackage{colordvi}  
\definecolor{Red}{rgb}{1,0,0}

\makeatletter
\def\authornote{\xdef\@thefnmark{$\dagger$}\@footnotetext}
\makeatother

\begin{document}

\title{Ultrafast polariton-phonon dynamics of strongly coupled quantum dot-nanocavity systems}

\author{Kai M\"uller$^\dagger$}
\email{kaim@stanford.edu}
\affiliation{E. L. Ginzton Laboratory, Stanford University, Stanford, California 94305, USA}
\author{Kevin A. Fischer$^\dagger$}
\authornote{These authors contributed equally.}
\affiliation{E. L. Ginzton Laboratory, Stanford University, Stanford, California 94305, USA}
\author{Armand Rundquist}
\affiliation{E. L. Ginzton Laboratory, Stanford University, Stanford, California 94305, USA}
\author{Constantin Dory}
\affiliation{E. L. Ginzton Laboratory, Stanford University, Stanford, California 94305, USA}
\author{Konstantinos G. Lagoudakis}
\affiliation{E. L. Ginzton Laboratory, Stanford University, Stanford, California 94305, USA}
\author{Tomas Sarmiento}
\affiliation{E. L. Ginzton Laboratory, Stanford University, Stanford, California 94305, USA}
\author{Yousif A. Kelaita }
\affiliation{E. L. Ginzton Laboratory, Stanford University, Stanford, California 94305, USA}
\author{Victoria Borish}
\affiliation{E. L. Ginzton Laboratory, Stanford University, Stanford, California 94305, USA}
\author{Jelena Vu\v{c}kovi\'c}
\affiliation{E. L. Ginzton Laboratory, Stanford University, Stanford, California 94305, USA}

\date{\today}

\begin{abstract}
We investigate the influence of exciton-phonon coupling on the dynamics of a strongly coupled quantum dot-photonic crystal cavity system and explore the effects of this interaction on different schemes for non-classical light generation. By performing time-resolved measurements, we map out the detuning-dependent polariton lifetime and extract the spectrum of the polariton-to-phonon coupling with unprecedented precision. Photon-blockade experiments for different pulse-length and detuning conditions (supported by quantum optical simulations) reveal that achieving high-fidelity photon blockade requires an intricate understanding of the phonons' influence on the system dynamics. Finally, we achieve direct coherent control of the polariton states of a strongly coupled system and demonstrate that their efficient coupling to phonons can be exploited for novel concepts in high-fidelity single photon generation.
\end{abstract}

\pacs{Valid PACS appear here}

\maketitle

\section{Introduction}
The strong coupling between a single photon and a single quantum emitter is of substantial interest for both investigations of the fundamentals of quantum optics as well as potential applications in optical computing, quantum metrology, and quantum cryptography \cite{2008_Faraon_Blockade, Volz2012, 2014_Laussy}. This universality is reflected by the diversity of associated experimental realizations, ranging from those in atomic physics \cite{Hood1998,Mckeever2003,Mabuchi2002} to superconducting systems \cite{2008_Fink, Niemczyk2010} and semiconductor devices \cite{2004_Reithmaier}. In the solid state, self-assembled quantum dots (QDs) are the most investigated quantum emitters due to their strong interaction with light as well as their nearly transform-limited linewidth \cite{Hogele2004, Prechtel2013, Kuhlmann2013}. Optical resonators, in the form of both micropillar structures \cite{2004_Reithmaier} and photonic crystal cavities \cite{2007_Hennessy}, are widely used to enhance this light-matter interaction. Photonic crystal cavities are especially promising for on-chip integration of quantum optical circuits due to the convenient fabrication of integrated waveguide and detector structures \cite{2013_Reithmaier}. In contrast to other systems, semiconductor quantum emitters are usually embedded in a crystalline host matrix, resulting in strong interactions with phonons. While in many circumstances such interactions are undesirable, it is also possible to carefully design experiments which benefit from the presence of phonons. For example, it was recently demonstrated that the QD coupling to phonons can be exploited for a robust and high-fidelity preparation of exciton and biexciton states \cite{Glassl2013, Pelle2014, Quilter2015}. Meanwhile, when QDs are coupled to semiconductor optical cavities, far off-resonant feeding effects have been shown to be consistent with a phonon-induced interaction \cite{Winger2009, Calic2011}. In fact, for QD-cavity systems the coupling to phonons and its impact on applications has been extensively studied in the weakly coupled regime \cite{Hohenester2009, Suff2009}. However, for strongly coupled systems these effects remain largely unexplored experimentally. As such systems become increasingly relevant for their potential in building quantum optical networks, it is essential to develop a solid understanding of the influence that phonons have on the system dynamics.

In this paper, we investigate the interaction of a strongly coupled QD-cavity system with phonons and explore the impact of this coupling on different schemes for non-classical light generation. By performing temporally resolved spectroscopy experiments, we map out the detuning-dependent polariton lifetime and extract the spectrum of the polariton-to-phonon coupling. As a result, we conclude that the high efficiency of this coupling entails its importance for all applications of strongly coupled solid-state systems. To corroborate this finding, we investigate the influence of electron-phonon interaction on one of the key applications of strongly coupled systems: non-classical light generation. To this end, we perform photon-blockade experiments and complementary quantum optical simulations under different pulse length and detuning conditions, which reveal that the pulse length has to be chosen correctly for a given detuning in order to obtain high-fidelity photon blockade. In particular, the impact of coupling to phonons is found to be more pronounced for the case of detuned photon blockade, which is the condition that yields the highest fidelity and efficiency \cite{Kai2014}. Finally, we demonstrate that the efficient coupling to phonons can also be utilized for novel concepts in high-fidelity single photon generation. 
\\
$ $\\
$ $\\

\section{Strongly coupled QD-cavity systems}
The sample investigated consists of a layer of low density InAs QDs grown by molecular beam epitaxy and embedded in a photonic crystal L3 cavity \cite{2003_Akahane}. The energy structure of a QD strongly coupled to an optical cavity (without phonon coupling) is usually described by the Jaynes-Cummings (JC) Hamiltonian
\begin{equation}
H = \omega_a a^{\dagger} a + (\omega_ a + \Delta)\sigma^{\dagger}\sigma + g(a^{\dagger}\sigma + a\sigma^{\dagger})
\label{equation:1}
\end{equation}
where $\omega_a$ denotes the frequency of the cavity,  $a$ the cavity mode operator, $\sigma$ the lowering operator of the quantum emitter, $\Delta$ the detuning between quantum emitter and cavity, and $g$ the emitter-cavity field coupling strength. After introducing dissipation into the JC system we obtain a quantum Liouville equation with complex eigenenergies \cite{2012_Laussy_Climbing}:
\begin{multline}
E^n_{\pm} = n\omega_a + \frac{\Delta}{2}-i\frac{(2n-1)\kappa + \gamma}{4} \\ \pm \sqrt{(\sqrt{n}g)^2+\left(-\frac{\Delta}{2}-i\frac{\kappa-\gamma}{4}\right)^2}
\label{equation:2}
\end{multline}
where $E^n_{\pm}$ corresponds to the $n$th rung of the system and $\kappa$ and $\gamma$ to the cavity and QD energy decay rates, respectively. The real parts of $E^n_{\pm}$ yield the energies of the states whereas the imaginary parts yield their linewidths. Note that all four parameters $\kappa$, $\gamma$, $\Delta$ and $g$ contribute to the polariton splittings and linewidths.

\begin{figure}[!t]
  \includegraphics[width=\columnwidth]{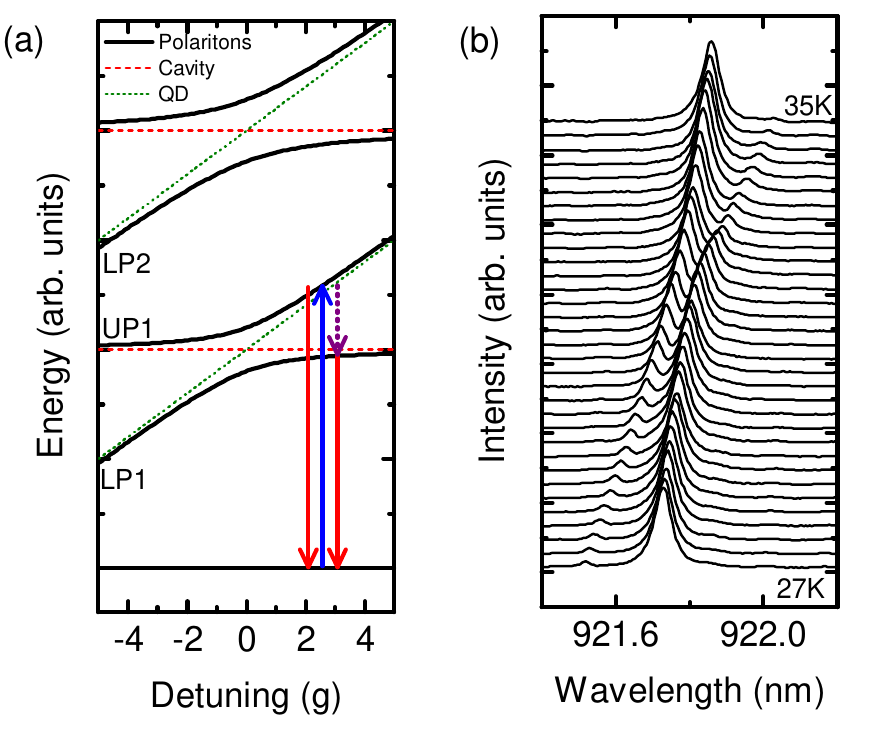}
  \caption{(a) Schematic illustration of the energy structure of a strongly coupled system as given by the Jaynes-Cummings ladder obtained from equation \ref{equation:2}. The blue arrow illustrates resonant excitation of UP1, the red arrows photon decay and the purple arrow phonon-assisted transfer. (b) Cross-polarized reflectivity spectra of the coupled QD-cavity system obtained by temperature-tuning the QD through the cavity resonance. An anticrossing of the peaks clearly demonstrates the strong coherent coupling.}
  \label{figure:1}
\end{figure}

The resulting eigenenergies, the Jaynes-Cummings-ladder dressed states, are presented in figure \ref{figure:1}a. They consist of a series of anticrossing branches that are labelled UP$n$ (LP$n$) for the upper (lower) polariton where $n$ is the index of the rung. For $n$ photons in the cavity the energy is $n \omega_a$ (dashed red lines), and the energy of the quantum emitter (dotted green line) varies with a detuning parameter. Due to the coupling, the resulting energy eigenstates are the anticrossing polariton branches. At resonance, the splitting is given by $2g\sqrt{n}$.

Experimentally, we can observe the splitting of the lowest-manifold polaritons in cross-polarized reflectivity \cite{2007_Englund} through control of the crystal lattice temperature. The result of such an experiment is presented in figure \ref{figure:1}b and shows a clear anticrossing. A fit to the data reveals values of $g = 12.3 \cdot 2\pi$ GHz and $\kappa = 18.4 \cdot 2\pi$ GHz. The radiative lifetime of a QD in a bulk photonic environment is known to be $\sim 1$ ns and thus, for QDs in photonic crystal cavities $\gamma$ is much smaller than $\kappa$. Due to the photonic bandgap $\gamma$ is further suppressed and reported values for the continuum mode lifetimes are around $4-12$ ns \cite{Englund2005, Kaniber2008} corresponding to $\gamma \sim 0.01 \cdot 2\pi$ GHz.

\section{Polariton decay rates}
\begin{figure}[!t]
  \includegraphics[width=\columnwidth]{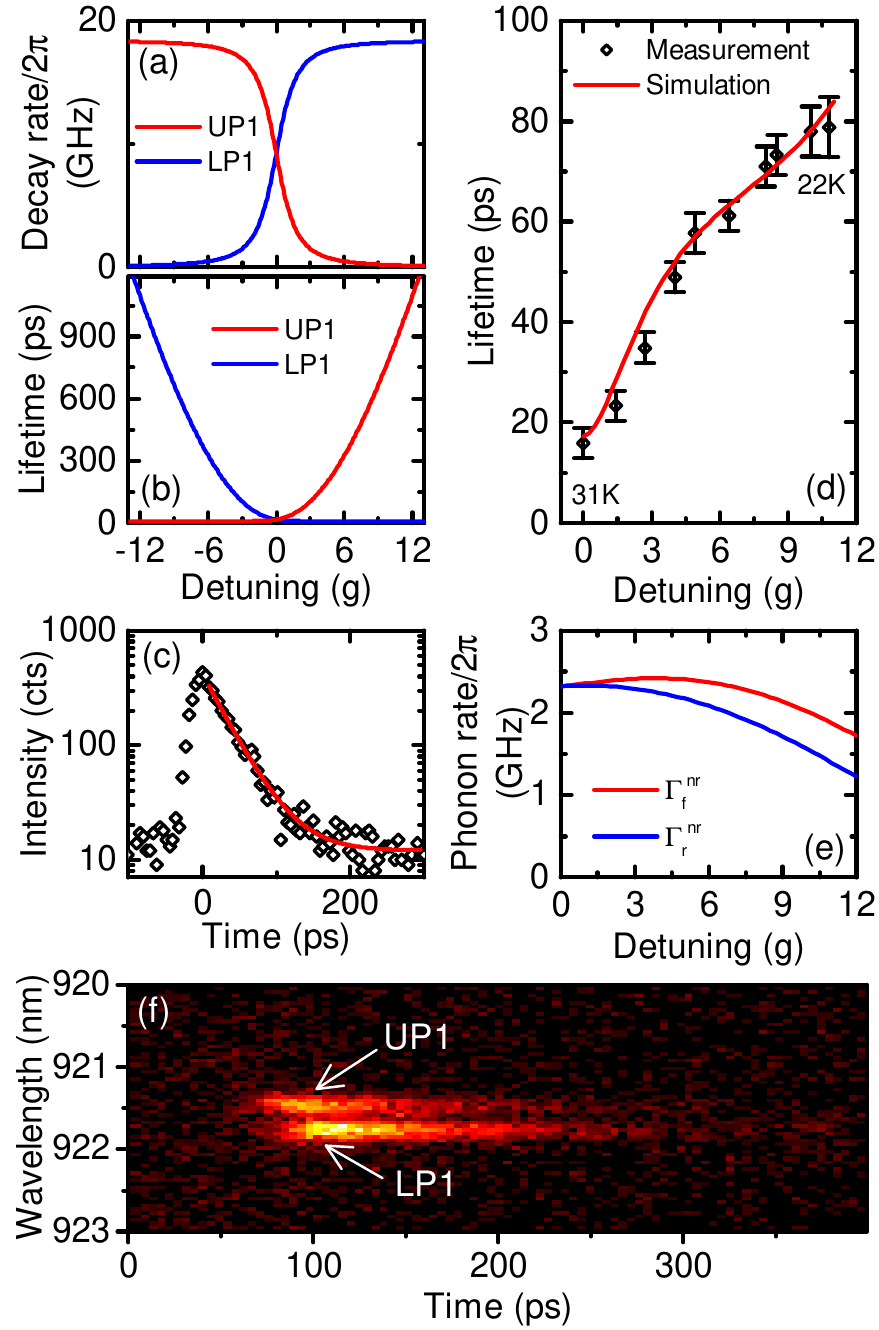}
  \caption{(a) Decay rates and (b) resulting lifetimes of the polariton branches as functions of the QD-cavity detuning, obtained from equation \ref{equation:2}. (c) Time-resolved spectrally integrated luminescence intensity at a detuning of $\Delta=2.8\,g$ for excitation of UP1. The lifetime is extracted from a mono-exponential fit (shown in red). (d) Measured and simulated lifetime of UP1 as a function of the detuning. (e) Phonon-assisted state transfer rates extracted from the lifetime at each detuning. (f) Spectrally and temporally resolved luminescence at a QD-cavity detuning of $\Delta=7\,g$. The delayed onset of emission from LP1 for resonant excitation of UP1 is consistent with phonon-assisted population transfer.}
  \label{figure:2}
\end{figure}

Using the values above and equation \ref{equation:2}, we plot the decay rates (imaginary parts of $E^n_{\pm}$) and radiative lifetimes of the polariton branches of an ideal Jaynes-Cummings system as a function of the detuning in figure \ref{figure:2}a and \ref{figure:2}b, respectively. In these plots, the red (blue) curve corresponds to UP1 (LP1), which is more cavity-like (QD-like) for negative and QD-like (cavity-like) for positive detunings. The decay rates and thus radiative lifetimes exchange over a range that is determined by $g$ and $\kappa$.

In contrast to ideal JC-systems, however, for solid-state systems a phonon-assisted population transfer between QD excitons and cavity photons is known to exist \cite{Hohenester2009, Hohenester2010, Hughes2011, Roy2011}. Thus, we sought to determine the true state lifetimes through time-resolved measurements. The fast cavity decay rate and strong interaction with phonons are expected to result in significantly shortened lifetimes. Hence, we performed the measurements using pulsed resonant excitation and time-resolved detection on a streak camera with sub-5 ps resolution. While non-resonant excitation would lead to slow carrier relaxation into the QDs \cite{Reithmaier2014} our resonant scheme minimizes timing jitter in the excitation. To avoid exciting higher rungs of the JC ladder, we pumped the more QD-like polariton branch UP1 (as schematically illustrated by the blue arrow on the right side of figure \ref{figure:1}a) with a pulse length that was carefully chosen to be shorter than the state lifetime but spectrally sharp enough not to overlap with the other polariton branch. Luminescence was then observed either from this state directly (red arrow in figure \ref{figure:1}a) or from the more cavity-like polariton branch after a phonon-assisted transfer (purple and red arrows in figure \ref{figure:1}a). The dynamics of this process can be simulated using a rate equation model with four rates: radiative recombination from UP1 ($\Gamma^r_{UP1}$), radiative recombination from LP1 ($\Gamma^r_{LP1}$), the phonon-assisted transfer rate from UP1 to LP1 ($\Gamma^{nr}_{f}$), and vice versa ($\Gamma^{nr}_{r}$). As shown in figure \ref{figure:2}c, the spectrally integrated photoluminescence intensity of a typical time-resolved measurement at a QD-cavity detuning of $\Delta = 2.8 \, g$ (including contributions from both decay paths) shows a mono-exponential decay, as expected from the rate equation model. A fit to the data (red line in figure \ref{figure:2}c) reveals a decay time of $34.9 \pm 2$ ps. We note here that the spectrally integrated intensity decays with the same time constant as the UP1 line (see supplemental material for details on the rate equation model).

To map out the polariton lifetimes we repeated the time-resolved measurement presented in figure \ref{figure:2}c for different detunings in the range $\Delta = 0-11 \, g$. As shown in figure \ref{figure:2}d, the resulting lifetime of UP1 increases with increasing detuning. However, the increase occurs much more slowly than indicated by the rates calculated from the JC model assuming no phonon interaction presented in figure \ref{figure:2}b. For the detunings investigated here, LP1 is more cavity-like and, thus, has a much shorter radiative lifetime, ranging from 16 ps down to 8 ps as the detuning is increased. Meanwhile, the JC lifetime without phonon interaction for UP1 varies from 16 ps up to almost 1000 ps, while the measured lifetime of UP1 only increases up to roughly 80 ps. Therefore, a strong phonon-assisted population transfer from UP1 to LP1 significantly shortens the observed lifetime of UP1.

We model this phonon-assisted population transfer using an effective master equation derived in a polaron frame with respect to the phonon interaction \cite{Roy2011}. This model fully captures the polariton dynamics and correlations discussed here without the need for non-Markovian damping. Additionally, the model is valid for our system over a wide range of temperatures, even extending to near absolute zero (see supplemental material for details on the model and its validity). A fit to the data including this model is presented as a red line in figure \ref{figure:2}d and produces very good overall agreement with all of the measured data. Since the detuning between cavity and QD was controlled by the lattice temperature, the relative temperature of each detuning was taken into account for these simulations. From the fit we can extract the phonon-assisted population transfer rates $\Gamma^{nr}_{f}$ and $\Gamma^{nr}_{r}$. They are presented in figure \ref{figure:2}e as red and blue lines, respectively. Near resonance the two rates are very close due to the elevated temperature of $T=31\, K$. With increasing detuning both rates decrease as the polariton states evolve towards the bare QD and cavity states, where the phonon interaction cannot couple polaritons through the cavity drive. The lower values of $\Gamma^{nr}_{r}$ compared to $\Gamma^{nr}_{f}$ result from the different probabilities of phonon emission and absorption at low temperatures.

In order to further corroborate this model of phonon-assisted population transfer we performed measurements with simultaneous spectral and temporal resolution. The result of a typical measurement recorded at $\Delta =7 \, g$ is presented in figure \ref{figure:2}f. Two distinct luminescence peaks are visible. Most importantly the onset of luminescence from LP1 is delayed with respect to the onset of luminescence from UP1, as population must be transferred before photons can be emitted. At this detuning, the radiative recombination rate $\Gamma^{r}_{UP1}$ is small compared to the strong phonon-assisted population transfer and subsequent radiative recombination of LP1 ($\Gamma^{r}_{LP1}$). Hence, the strongest luminescence intensity is observed from LP1 even though UP1 is resonantly excited (for details see supplemental material).

\section{Pulse-length dependent photon blockade}
Having mapped out the detuning-dependent polariton lifetimes of the strongly coupled QD-cavity system we can now apply this knowledge to applications where the lifetime is critical. One of the most remarkable applications of strongly coupled systems is photon blockade \cite{2008_Faraon_Blockade, 2011_Reinhard_Correlated}, in which a laser pulse is tuned to the first rung of the JC-ladder and (due to the anharmonicity) is not resonant with higher transitions up the ladder. The JC system then transmits the portion of the incident pulse that couples to the system, which we observe in our cross-polarized reflectivity measurements. Therefore, the probability of coupling a single photon through the system is enhanced over that of multi-photon states, which results in a transmitted light beam that has a strong non-classical character. The fidelity of this process for the generation of single photons is inherently limited due to the linewidth of the polariton branches. Nevertheless, it was recently shown that in a strongly coupled system that is detuned by a few $g$, not only the purity but also the efficiency of single photon generation increases \cite{Kai2014}. As we will see, photon blockade depends crucially on the polariton lifetimes, especially in the detuned case.

\begin{figure}[!t]
  \includegraphics[width=\columnwidth]{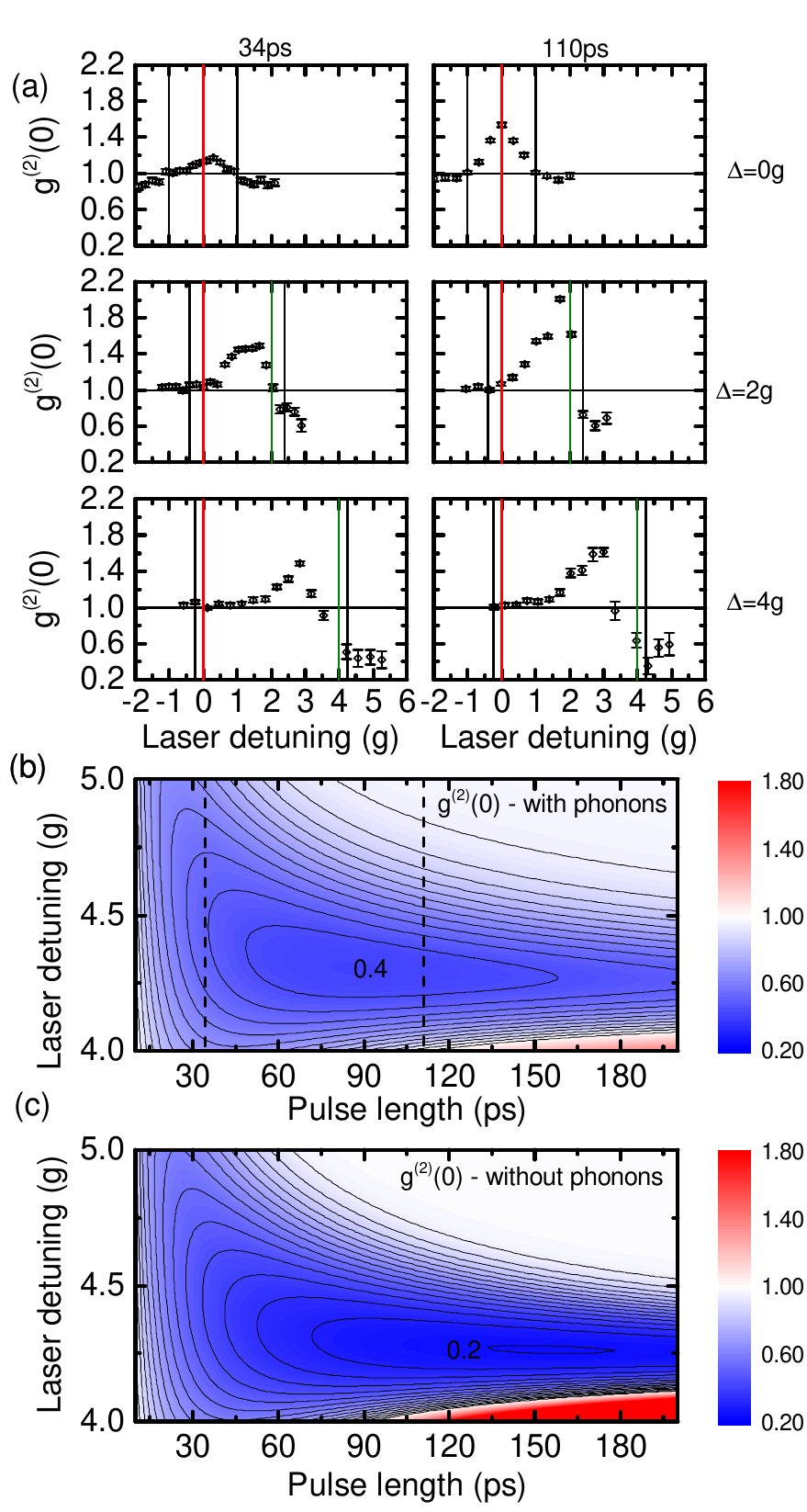}
  \caption{(a) Measured second-order coherence $g^{(2)}(0)$ as a function of the laser detuning for different QD-cavity detunings (top to bottom) and excitation pulse lengths (left to right). In each case, the red vertical line indicates the bare cavity frequency, the green vertical line indicates the bare QD frequency, and the black vertical lines indicate the polariton frequencies. (b-c) Simulated photon-blockade values of $g^{(2)}(0)$ as a function of the laser detuning and pulse length including (b) and excluding (c) the effect of phonons, both at a QD-cavity detuning of $\Delta = 4\,g$. To aid in comparison with the experimental results in (a), the dashed lines in (b) indicate the simulated anti-bunching values for photon blockade at the experimental pulse lengths.}
  \label{figure:3}
\end{figure}

To quantify the non-classical character of the photocount distribution, we use its measured degree of second-order coherence for zero time delay \cite{1963_Glauber, 2000_Loudon}
\begin{equation}
g^{(2)}(0) = \frac{\left\langle n(n-1) \right\rangle}{\left\langle n \right\rangle^2}
\label{equation:3}
\end{equation}
computed from expectations of the measured photocount distribution, where $n$ signifies a number of detections. Only non-classical light sources may have a second-order coherence $g^{(2)}(0)<1$, and $g^{(2)}(0)=0$ is measured exclusively for pulses with single-photon-like character. Owing to a cavity lifetime much shorter than the timing jitter of the single photon counters, measurements of $g^{(2)}(0)$ on our system can only be performed under pulsed excitation; this is in any case the experimental configuration required for on-demand applications. Hence, the choice of pulse length forces a compromise between frequency resolution (reducing the overlap of different rungs) and the likelihood of re-excitation of the system. In other words, if the laser pulses are too long the system will be re-excited during the interaction with a single pulse, reducing the non-classical character of the transmitted light. On the other hand, pulses with a shorter duration are spectrally broader, resulting in a larger overlap with higher rungs. With increasing detuning between QD and cavity resonances, the lifetime of the emitter-like (cavity-like) polariton branch increases (decreases). Therefore, in order to obtain the strongest photon blockade the pulse length has to be chosen according to the detuning-dependent polariton lifetime.

To test this hypothesis we performed measurements of $g^{(2)}(0)$ using a Hanbury Brown and Twiss (HBT) type experiment \cite{2001_Santori} for different QD-cavity detunings (and thus polariton lifetimes) and pulse lengths. Figure \ref{figure:3}a shows $g^{(2)}(0)$ as a function of the laser detuning for pulse lengths \footnote{All pulse lengths are quoted as full-width half-maximum throughout the text.} of $34$ ps on the left and $110$ ps on the right, and detunings of $\Delta = 0 \, g$, $\Delta = 2 \, g$ and $\Delta = 4 \, g$ from top to bottom. At $\Delta = 0$ both pulse lengths result in a symmetric curve with photon-blockade dips of $g^{(2)}(0)<1$ for laser detunings of $1 - 2 \,g$. We measure a minimum $g^{(2)}(0) = 0.88 \pm 0.03$ for $34$ ps pulses and $g^{(2)}(0) = 0.93 \pm 0.03$ for 110 ps pulses. Due to the short polariton lifetime of $16$ ps at $\Delta =0$ the longer pulses result in an increased likelihood of re-excitation during the presence of a single pulse, which leads to a higher value of $g^{(2)}(0)$ in photon blockade. We note that since the radiative recombination rates are very fast when the QD and the cavity are in resonance, phonons have only a minor effect in this case (see supplemental material for simulations and details).

With increasing detuning the fidelity of photon blockade increases as expected \cite{Kai2014}, and for $\Delta =4 \, g$ (bottom plots of figure \ref{figure:3}a) we measure values as low as $g^{(2)}(0) = 0.44 \pm 0.07$ for 34 ps pulses and $g^{(2)}(0) = 0.34 \pm 0.07$ for 110 ps pulses. In particular, due to the longer polariton lifetime at $\Delta = 4 \, g$, we observe a lower value of $g^{(2)}(0)$ for the longer pulses. Here, re-excitation limits the minimum of the second-order coherence for the long pulses, while spectral overlap with transitions to higher rungs limits the minimum for the short pulses.

To explain this finding we performed quantum optical simulations using the Quantum Optics Toolbox in Python (QuTiP) \cite{2014_Johansson} based on the Quantum Regression Theorem \cite{Kai2014}. The results of these simulations are presented in figure \ref{figure:3}b-c. The figures show $g^{(2)}(0)$ as a function of the laser detuning and pulse length for $\Delta = 4 \, g$, both including (figure \ref{figure:3}b) and excluding (figure \ref{figure:3}c) phonon-assisted population transfer (see supplemental material for details on the simulations and additional representations). In both cases, with increasing pulse length the photon-blockade dip narrows down (as seen in the experiment) due to the improved spectral resolution of longer pulses. Moreover, the minimum value of $g^{(2)}(0)$ decreases due to the decreased overlap with higher rungs (as well as the other polaritonic branch) before increasing due to re-excitation for too-long pulses. As can be seen by comparing figure \ref{figure:3}b with figure \ref{figure:3}c, the phonon-induced reduction of the polariton lifetime means that the deepest photon-blockade dip is observed at a shorter pulse length when phonons are included, and the corresponding minimum $g^{(2)}(0)$ values are a little higher. The simulations are in very good overall agreement with the measurements presented in figure \ref{figure:3}a, and they highlight the importance of taking phonon-assisted transfer into account when selecting the optimal pulse length.

For an excitation between the polariton branches, a peak with $g^{(2)}(0)>1$ is observed for all detunings and pulse lengths. This phenomenon is known as photon-induced tunneling \cite{2008_Faraon_Blockade}, where anharmonicity of the JC ladder results in bunching. However, the bunching values peak at the position where the probability of zero detections is highest and simulations of the photodetection distributions barely deviate from those of a coherent state with a similar count rate \cite{Kai2014}. Therefore, in this report we focus on the antibunching regime.

\section{Coherently-excited single-photon source}
\begin{figure}[!t]
  \includegraphics[width=\columnwidth]{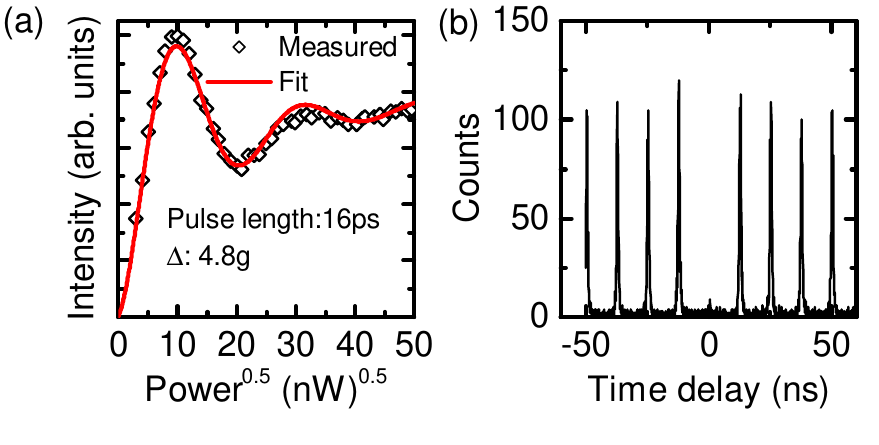}
  \caption{(a) Rabi oscillation observed upon exciting UP1 and detecting LP1. (b) Measurement of the second-order coherence for exciting UP1 with a $\pi$ pulse and detecting LP1 yielding $g^{(2)}(0)=0.03 \pm 0.01$}
  \label{figure:4}
\end{figure}

While we have seen above that for detuned photon blockade the presence of exciton-phonon coupling slightly decreases the fidelity, this coupling also allows us to investigate more sophisticated schemes for on-demand single-photon generation. In particular, the population of UP1 can be coherently controlled if the pulse length is chosen shorter than the state lifetime but spectrally narrow enough to avoid exciting higher transitions up the JC ladder. Due to coherent scattering of the laser and imperfect suppression of the laser reflection from the sample surface, it is difficult to observe Rabi oscillations directly. However, phonon-assisted emission from LP1 following excitation of UP1 occurs at a different frequency. In addition, for detunings in the range of $\Delta = 5-10 \, g$, $\Gamma^{r}_{UP1}$ is strongly Purcell-suppressed while $\Gamma^{r}_{LP1}$ is very fast. Most strikingly, $\Gamma^{nr}_{f}$ is proportional to $g^2$ \cite{Roy2011} and, thus, very efficient for a strongly coupled system. Therefore, even when resonantly exciting UP1 for these detunings most luminescence is emitted from LP1. The result of such an experiment is presented in figure \ref{figure:4}a, which shows the emission intensity from LP1 for a resonant excitation of UP1 with 16 ps pulses at a detuning of $\Delta = 4.8 \, g$. Clear Rabi oscillations are observed and fitted (red line in figure \ref{figure:4}a) using a damped oscillation superimposed with a very weak background that is linear with the excitation power, due to a finite leakage of the excitation laser into the detection channel. This power-dependent damping is a direct consequence of the polariton-phonon interaction. With increasing laser power the polariton levels hybridize, mixing higher rungs with UP1. This increases the number of available damping channels via the polariton-phonon coupling and results in a damped oscillation.

To investigate single photon generation in this configuration we measure $g^{(2)}(0)$ while exciting UP1 with a $\pi$ pulse and detecting LP1 emission. The result of this experiment is presented in figure \ref{figure:4}b and shows almost perfect antibunching. By integrating the area of the peaks we obtain a value of $g^{(2)}(0)=0.04 \pm 0.01$. Fitting the data with a series of Gaussian peaks is a little less sensitive to noise from the dark counts of the detectors and yields $g^{(2)}(0)=0.03 \pm 0.01$ (see supplemental material for the fit and a closer examination of the data). These results rival the best values obtained from QDs, while gaining advantages provided by the optical cavity (e.g., the very short lifetime that enables a high photon generation rate and the potential for integration into on-chip optical networks). Hence, our findings clearly demonstrate the potential of exploiting the efficient exciton-phonon coupling of strongly coupled QD-cavity systems for single photon generation.

\section{Summary and conclusions}
The presence of phonons represents one of the key distinguishing features of the solid state. Our investigations have laid the groundwork for understanding their role in the dynamics of strongly coupled systems. We have mapped out the detuning-dependent polariton lifetime of a strongly coupled QD-cavity system and extracted the spectrum of the polariton-to-phonon coupling for QD-cavity detunings up to $11 \, g$. We have shown that in order to obtain high-fidelity photon blockade a pulse length must be chosen that depends on the QD-cavity detuning. Finally, we have achieved direct coherent control of polariton states and provided an unprecedented demonstration of how the efficient coupling to phonons can be exploited for high-fidelity single photon generation. Our findings suggest that the influence of the efficient coupling to phonons should be considered for all applications of strongly coupled solid-state systems. Furthermore, we expect that our contribution towards understanding strongly coupled systems in the solid state will play a vital role in utilizing QD-cavity platforms for novel physics such as higher-order non-classical light generation \cite{2014_Laussy}. 

We gratefully acknowledge financial support from the Air Force Office of Scientific Research, MURI center for multifunctional light-matter interfaces based on atoms and solids (Grant No. FA9550-12-1-0025)”  and support from the Army Research Office (Grand No. W911NF1310309). KM acknowledges support from the Alexander von Humboldt Foundation. KAF acknowledges support from the Lu Stanford Graduate Fellowship and the National Defense Science and Engineering Graduate Fellowship. KGL acknowledges support from the Swiss National Science Foundation. VB acknowledges support from the National Science Foundation. YAK acknowledges support from the Stanford Graduate Fellowship and the National Defense Science and Engineering Graduate Fellowship.


\providecommand{\noopsort}[1]{}\providecommand{\singleletter}[1]{#1}%

\cleardoublepage

\onecolumngrid
\section*{Supplemental Material}
$ $\\
$ $\\
$ $\\
\twocolumngrid

\section*{Methods}
\textbf{Sample fabrication}\\
The MBE grown structure consists of a $\sim 900 \, \text{nm}$ thick $Al_{0.8}Ga_{0.2}As$ sacrificial layer followed by a $145 \, \text{nm}$ thick GaAs layer that contains a single layer of InAs QDs. Our growth conditions result in a typical QD density of $60-80 \, \mu \text{m}^{-2}$. The photonic crystals were fabricated using $100 \, \text{keV}$ e-beam lithography with ZEP resist, followed by reactive ion etching and HF removal of the sacrificial layer. The photonic crystal lattice constant was $a = 246 \, \text{nm}$ and the hole radius $r \approx 60 \, \text{nm}$. The cavity fabricated is a linear three-hole defect (L3) cavity. To improve the cavity quality factor, holes adjacent to the cavity were shifted \cite{2005_Akahane, 2014_Minkov}.
\\

\textbf{Optical spectroscopy}\\
All optical measurements were performed with a liquid helium flow cryostat at temperatures in the range $10-40 \, \text{K}$. For excitation and detection a microscope objective with a numeric aperture of $NA = 0.75$ was used. Cross-polarised measurements were performed using a polarising beam splitter. To further enhance the extinction ratio, additional thin film linear polarisers were placed in the excitation/detection pathways and a single mode fibre was used to spatially filter the detection signal. Furthermore, two waveplates were placed between the beamsplitter and microscope objective: a half-wave plate to rotate the polarisation relative to the cavity and a quarter-wave plate to correct for birefringence of the optics and sample itself.
\\

\textbf{Autocorrelation measurements}\\
Second-order autocorrelation measurements were performed using a Hanbury Brown and Twiss (HBT) setup consisting of one fibre beamsplitter and two single photon avalanche diodes. The  detected photons were correlated with a PicoHarp300 time counting module.\\

\textbf{Simulations}\\
The simulations of $g^2(0)$ presented in figure 3 of the main text were performed using the Quantum Optics Toolbox in Python (QuTiP) \cite{2014_Johansson}. Here, we used a Jaynes-Cummings Hamiltonian evolution subject to dissipation terms of the Lindblad form. These terms include those due to the cavity decay and the phonon-induced cavity$\rightarrow$exciton and exciton$\rightarrow$cavity processes, with rates derived from the phonon model discussed below. The simulations are time-dependent, in that the cavity is driven with a Gaussian pulse of experimentally extracted pulse-width. The only free parameter is the laser power. Finally, the calculation of the pulse-wise second order coherence is exact, in that it is the sum of all two-time photon-photon correlations of cavity photons. These correlations are calculated via the quantum regression theorem \cite{Kai2014}. \\

\section*{Rate equation model for describing the system dynamics}
\begin{figure}[t!]
  \includegraphics[width=\columnwidth]{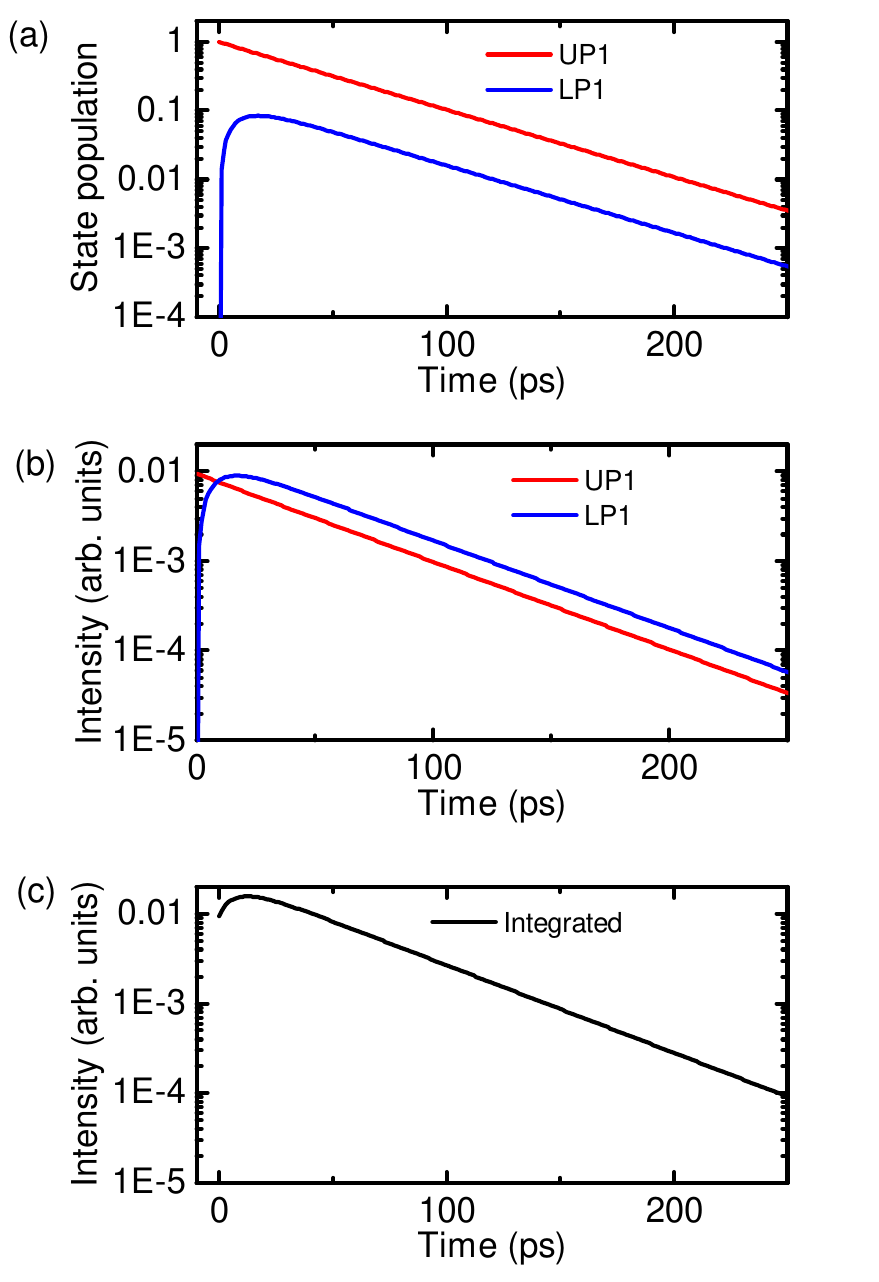}
  \caption{(a) Calculated dynamics of the state population of UP1 and LP1 after an excitation of UP1 at $\Delta = 3g$. (b) Calculated dynamics of the emission intensity from UP1 and LP1. (c) Calculated dynamics of the spectrally integrated emission intensity.}
  \label{figure:S1}
\end{figure}

As discussed in the main text, we use a rate equation model with two states and four rates to describe the system dynamics: radiative recombination from UP1 ($\Gamma^{r}_{UP1}$), radiative recombination from LP1 ($\Gamma^{r}_{LP1}$), the phonon-assisted transfer rate from the UP1 to LP1 ($\Gamma^{nr}_{f}$), and vice versa ($\Gamma^{nr}_{r}$). The resulting rate equation model then reads:
\begin{multline}
\frac{d}{dt}\left( \begin{matrix} P_{UP1}(t) \\ P_{LP1}(t)  \end{matrix} \right) = \\
\left( \begin{matrix} -(\Gamma^{r}_{UP1} + \Gamma^{nr}_{f}) & \Gamma^{nr}_{r} \\ \Gamma^{nr}_{f} & -(\Gamma^{r}_{LP1} + \Gamma^{nr}_{r}) \end{matrix} \right) \cdot
\left( \begin{matrix} P_{UP1}(t) \\ P_{LP1}(t)  \end{matrix} \right)
\end{multline}

where $P_{UP1}(t)$ and $P_{LP1}(t)$ describe the populations of the states UP1 and LP1, respectively. From solving these equations we obtain the time dependence of the state populations.

An example is presented in figure \ref{figure:S1}a that shows the populations of UP1 and LP1 at a detuning of $\Delta = 3g$ in red and blue for excitation of UP1 and using the values for the four rates as obtained in the main text for this detuning. After a short time the populations quasi-thermalize and decay with an almost perfect exponential shape. The resulting intensity of emission from these two states is presented in figure \ref{figure:S1}b using the same colour coding as figure \ref{figure:S1}a and shows the same behaviour. The spectrally integrated emission intensity is presented in figure \ref{figure:S1}c. Importantly, after the fast quasi-thermalization the emission intensities from the two states (figure \ref{figure:S1}b) decay with the same time constant. Therefore, the spectrally integrated intensity (figure \ref{figure:S1}c) decays with the same time constant. This allows extraction of the lifetime of UP1 from temporally analysing the spectrally integrated luminescence intensity. In particular, the streak camera that was used to record the temporally resolved spectra has a lower spectral resolution than the spectrometer and CCD that was used to record the time integrated spectra as presented in figure 1 of the main text. Therefore, for detunings smaller than $\Delta \approx 5 \, g$ analysing the spectrally integrated intensity is the only option as the two time traces cannot be spectrally resolved. We note here that these finding hold for all investigated sets of parameters.

\section*{Phonon model}
To model the polariton-phonon coupling we used a model that utilizes an effective master equation derived in a polaron frame with respect to the phonon interaction (also known as the effective phonon master equation, or EPME) \cite{Roy2011}. In this model a characteristic phonon spectral function is used which is given by
\begin{equation}
J(\omega)=\alpha_p \omega^3 exp\left( \frac{\omega^2}{2\omega_b^2} \right)
\end{equation}
and describes the interaction between electrons and the logitudinal acoustic phonons via deformation potential coupling. This function uses two parameters $\alpha_p$ and $\omega_b$. The latter one corresponds to a high frequency cutoff due to the electronic localization length. The fit of the detuning dependent polariton lifetime presented in figure 2 of the main text resulted in the values $\alpha_p = 3 ps^2$ and $\omega_b = 0.22$ meV. However, we note that these values vary quite a lot throughout literature. This can be understood from the fact that $J(\omega)$ is derived for a spherical quantum dot with similar extents for electron and hole wavefunctions. Moreover, experiments typically cover only a limited frequency range and therefore can be fitted with a range of parameters.

We now discuss the validity of our application of the effective phonon master equation. Given that the EPME results from a truncation of the cavity-exciton transfer process to first order, it is incapable of modelling the asymmetry in the forward and reverse rates that physically arises due to spontaneous phonon emission. However, at our experimental bath temperatures the phonon occupancies are elevated such that the EPME only incorrectly estimates these transfer rates by a few percent - well within experimental error. Furthermore, even at low temperatures the EPME is capable of accurately capturing the dynamics of a typical QD-cavity system. Here, the radiative decay rate dominates the state decay near zero detuning, where the asymmetry in the phonon-induced transfer rates is the largest. Away from zero detuning the EPME correctly captures the influence of spontaneous phonon emission because the lowest-order polaritonic states have a clear excitonic or photonic character that can be matched to the transfer terms. Therefore, we are justified in ignoring these small inaccuracies and modeling our system dynamics with the EPME. We also note here that spectral diffusion does not affect the ultrafast dynamics described here as it occurs on a much slower timescale \cite{Kuhlmann2013}.

\newpage

\section*{Photon blockade simulations}
\begin{figure}[!t]
  \includegraphics[width=\columnwidth]{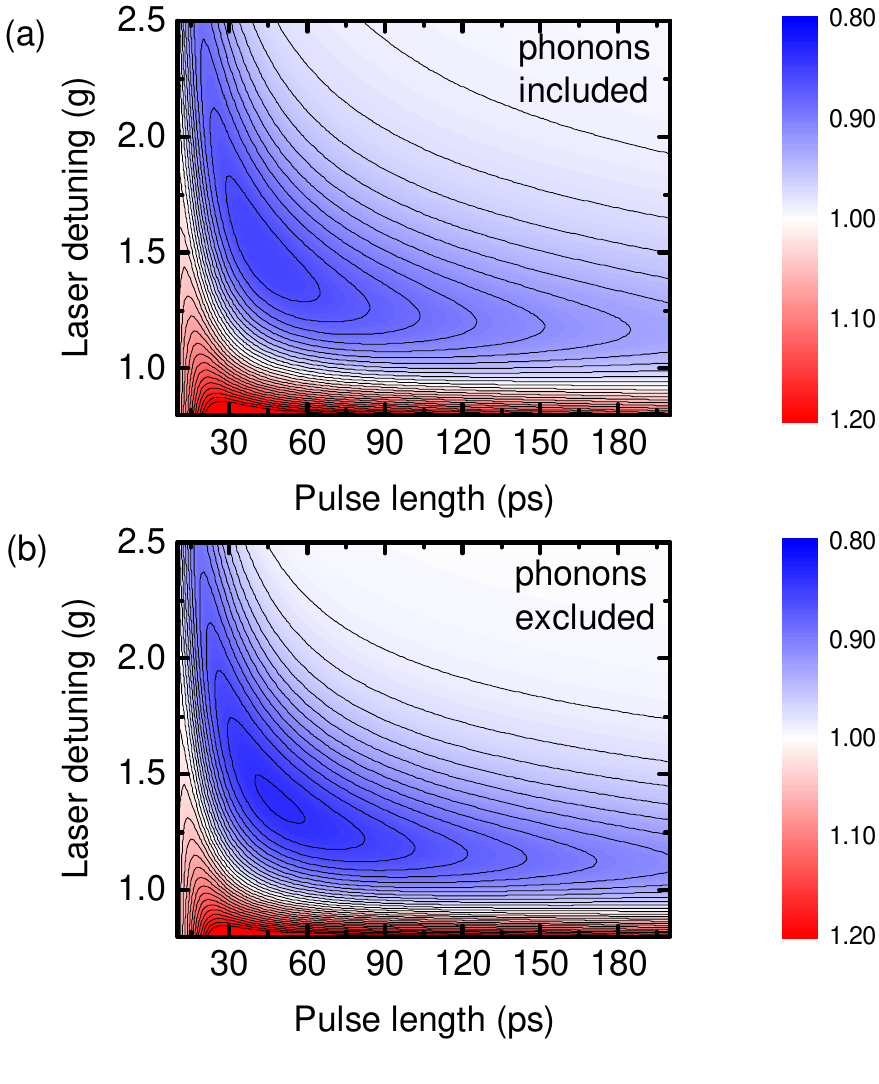}
  \caption{Simulated $g^2(0)$ as a function of the laser detuning and pulse length for a QD-cavity detuning of $\Delta = 0$ (a) including and (b) excluding phonons}
  \label{figure:S2}
\end{figure}

In the main text we presented simulations of $g^2(0)$ as a function of the pulse length and laser detuning for $\Delta = 4 \, g$ with and without phonons. As mentioned there, we also performed similar simulations for $\Delta = 0 \, g$. The result of these simulations is presented in figure \ref{figure:S2}a with phonons and figure \ref{figure:S2}b without phonons. In both cases the best photon-blockade is observed for relatively short pulses around $40 \, ps$ which can be understood from the short polariton lifetime at resonance which results in strong re-excitation of the system for longer pulses. Moreover, compared to the situation at $\Delta = 4 \, g$ (main text figure 3b and 3c) the difference in the optimum pulse length as well as observed value of $g^2(0)$ for the cases with and without phonons is much smaller. This results from the fact that the radiative decay rate of the polariton branches at resonance is much faster than the phonon mediated population transfer. Thus, the effect of phonons is much less pronounced. Regardless, since photon-blockade has a much higher efficiency at a detuning of a few $g$ compared to the resonant case, the impact of phonons on the polariton lifetime has to be considered carefully in order to optimize the experimental conditions for single photon generation by photon blockade.

\begin{figure}[!ht]
  \includegraphics[width=\columnwidth]{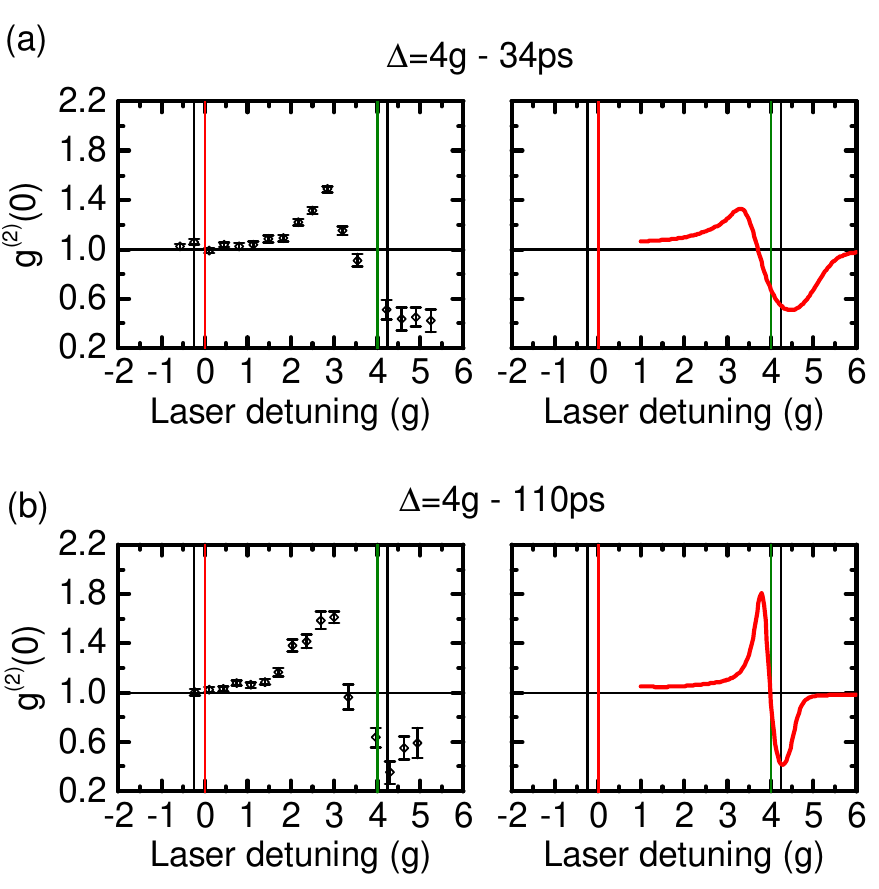}
  \caption{Comparison of measured and simulated $g^2(0)$ as functions of the laser detuning for a  QD-cavity detuning of $\Delta = 4 \, g$ with (a) 34 ps and (b) 110 ps long pulses.}
  \label{figure:S3}
\end{figure}

We also present a direct comparison between the measurements and simulations of $g^{(2)}(0)$ from figure 3 of the main text. To this end, we replotted in figure \ref{figure:S3} the measurements for $\Delta = 4g$ (from figure 3a of the main text) adjacent to simulations (from figure 3b of the main text) using the same representation. While there is some jitter in the $x$-values of the measurements due to the limited resolution of our frequency filter and temperature drift during the measurements, the simulations reproduce the experiments very well. In particular, the photon-blockade dip is quantitatively well reproduced. The simulations also reproduce the experimental trend that the bunching peak is higher for longer pulses. However, the bunching values should be loosely interpreted given that they occur at the position where the probability of zero detections is highest (as discussed in the main text). Thus, the bunching values are strongly altered by effects such as blinking of the QD or an imperfect suppression of the excitation laser.

\section*{Phonon-assisted single photon generation}
\begin{figure}[!t]
  \includegraphics[width=\columnwidth]{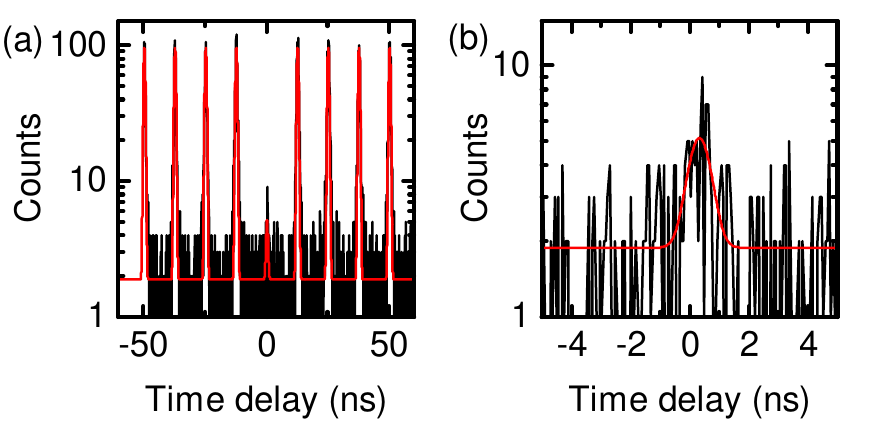}
  \caption{(a) Semi-logarithmic representation of the phonon-assisted single photon generation presented in figure 4b of the main text. A fit with a series of peaks is presented as a red line (b) Zoom on the center peak.}
  \label{figure:S4}
\end{figure}
In the main text, we presented the measurement of $g^{(2)}(0)$ for phonon-assisted single photon generation on a linear scale. Here, we represent the data on a semi-logarithmic scale in figure \ref{figure:S4}a to further visualise the single photon purity. A fit to the data with a series of peaks is presented as a red line and a magnification of the center peak is presented in figure \ref{figure:S4}b. Due to the short polariton lifetime compared to the repetition rate, a flat background resulting only from dark counts can be seen. This can be used to fit the average number of dark count coincidences, which due to the high temporal resolution of our measurement is only 1.88.

\end{document}